\documentclass{PoS}

\usepackage{setspace}

\def\hess{H.E.S.S.}
\def\hessii{H.E.S.S.-II}
\def\deg{$^{\circ}$}
\def\gr{$\gamma$-ray}
\def\grs{$\gamma$~rays}
\def\MSSG{\emph{MSSG}}
\def\pks{~\mbox{PKS 2155-304}}
\def\IRFS{\emph{IRFs}}

\title{Enhanced \hessii\ low energies performance thanks to the focus system}

\ShortTitle{The \hess-II focus system}

\author{\speaker{C.~Trichard}\\
LAPP, Université Savoie Mont Blanc, CNRS/IN2P3, Annecy-le-Vieux, France\\
        E-mail: \email{trichard@lapp.in2p3.fr}}
\author{A.~Fiasson\\
LAPP, Université Savoie Mont Blanc, CNRS/IN2P3, Annecy-le-Vieux, France\\
        E-mail: \email{fiasson@lapp.in2p3.fr}}
\author{G.~Maurin\\
LAPP, Université Savoie Mont Blanc, CNRS/IN2P3, Annecy-le-Vieux, France\\
        E-mail: \email{maurin@lapp.in2p3.fr}}
\author{G.~Lamanna\\
LAPP, Université Savoie Mont Blanc, CNRS/IN2P3, Annecy-le-Vieux, France\\
        E-mail: \email{lamanna@lapp.in2p3.fr}}
\author{on behalf the \hess\ Collaboration}

\abstract{
For the current generation of Imaging Atmospheric Cherenkov Telescopes (IACTs), with their large mirrors and their cameras with fine segmentation of photodetectors, the focusing capability is a relevant issue. The optical system of an IACT has a limited depth of field. Therefore, focusing the telescopes close to the shower maximum in the atmosphere has a significant impact on the data acquisition and analysis. As the distance of the shower maximum to the telescope depends (among others) on the zenith angle, an adjustable focus would be desirable. The fifth Cherenkov telescope of the \hessii\ array is equipped with a focus system which allows to adjust the position of the camera along the optical axis, possibly during data taking. This impact has been studied on gamma-ray Monte Carlo simulations, and the results in terms of gamma-ray trigger rate, energy reconstruction and gamma-ray effective area will be shown.
}

\FullConference{The 34th International Cosmic Ray Conference,\\
		30 July- 6 August, 2015\\
		The Hague, The Netherlands}

\begin{document}

\section{Introduction}

The fifth telescope of the \hess\ array is an unique instrument for many aspects. A focusing system has been installed on the camera pad. The distance of the camera to the dish can be adjusted to change the focus of the telescope. This feature was motivated by the increased focal length of this telescope compared to the smaller ones of the \hess\ array, that leads to a shorter depth of field. In this note, the performances granted by the focus system are presented. Comparisons between Monte Carlo simulations and data are provided.

\section{Hardware description}

The design and construction of the focus system for the fifth telescope has been challenging. Indeed, given the size of the telescope, the construction of a shelter enveloping the camera in telescope parking position was not possible. Moreover, the access to the camera for maintenance operation is very difficult, given the telescope structure and the inclination of the telescope in parking position. It was thus decided to construct a device to unload the camera. The focus system for the camera had to be compatible with such a device.

This focus system was designed to manage two main required functions: position the camera accurately on the focal axis and safely lock the camera on the telescope structure with an accuracy of 0.35 mm. The camera is locked in using a pneumatic system composed by four toggle fasteners and four jacks. For safety reasons, the control of this system is conditioned by hardware conditions.  A mobile part allows the displacement of the camera along the optical by the means of 4 rails and two independent brushless motors. The system has been initially positioned such that the movement range translates into a freedom of 156~mm (70mm)  away (toward) the dish from the position corresponding to the lid surface at the dish focal distance. 

\begin{figure}[ht!]
\centering
\includegraphics[width=0.4\textwidth]{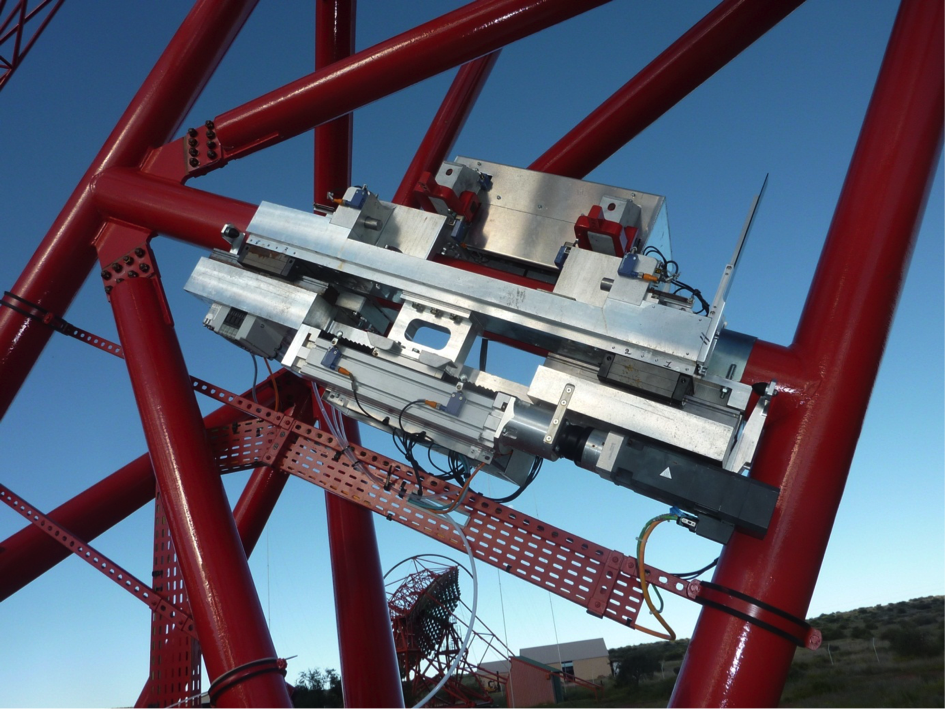}
\caption{Photography of the focus system in CT5 structure. The red tubular structure corresponds to the telescope structure. The grey part is the focus unit.}
\label{Hardware1}
\end{figure}

The system is fully automatized and is based on two parallel axes which are not mechanically coupled. The control architecture is based on a Programmable Logic Controller (PLC) which is composed by two Central Processing Unit, two fieldbus (one being synchronized) for the deported devices and specific variators which are able to manage power cuts and synchronized displacements. The monitoring of the system (coupled with the loading system) is done through an OPC server communicating with the PLC.  

\section{Monte Carlo simulations}

A previous study showed the benefits of the focus in term of trigger rate and reconstruction performance~\cite{Kray2013}. A simple Hillas reconstruction was used because of lack of advanced reconstructed method at that time. The available {\it model} analysis~\cite{ModelMono}, allow us to understand, quantify and optimize the effects of the focus with the best performance analysis in the \hess\ collaboration for the mono telescope analysis mode.

\subsection{Focus impact on the \gr\ images}

\begin{figure}
\centering
\includegraphics[width=0.6\textwidth]{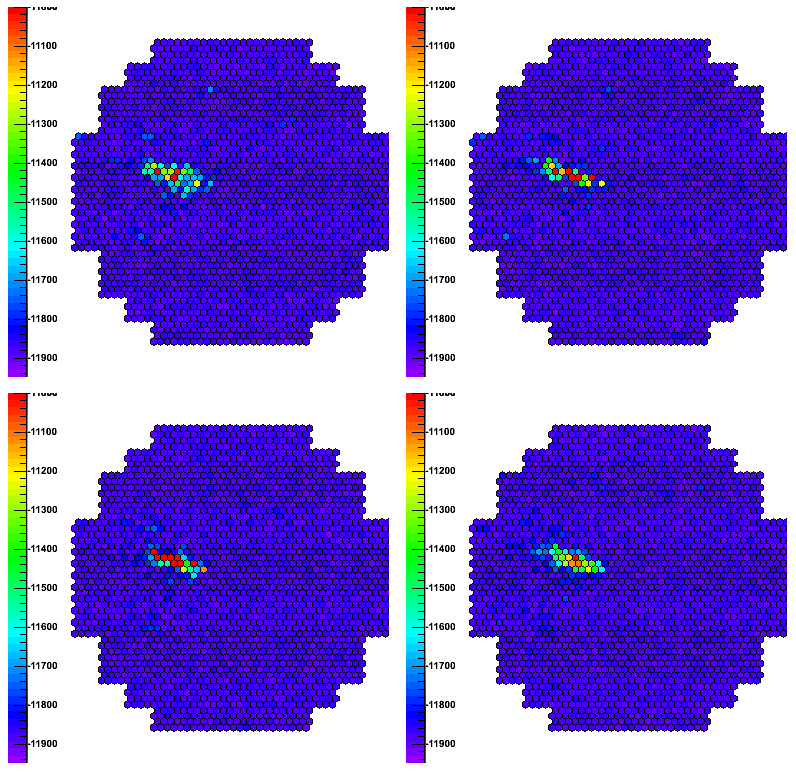} 
\caption{Left: Comparison of a 50~GeV simulated \gr\ for different focused distance. The telescope is focusing at 5~km in the upper left panel, at 10~km in the upper right panel, at 15~km in the lower left panel and at infinity in the lower right panel.}
\label{EventDisplay50}
\end{figure}

\begin{figure}
\centering 
\includegraphics[width=0.6\textwidth]{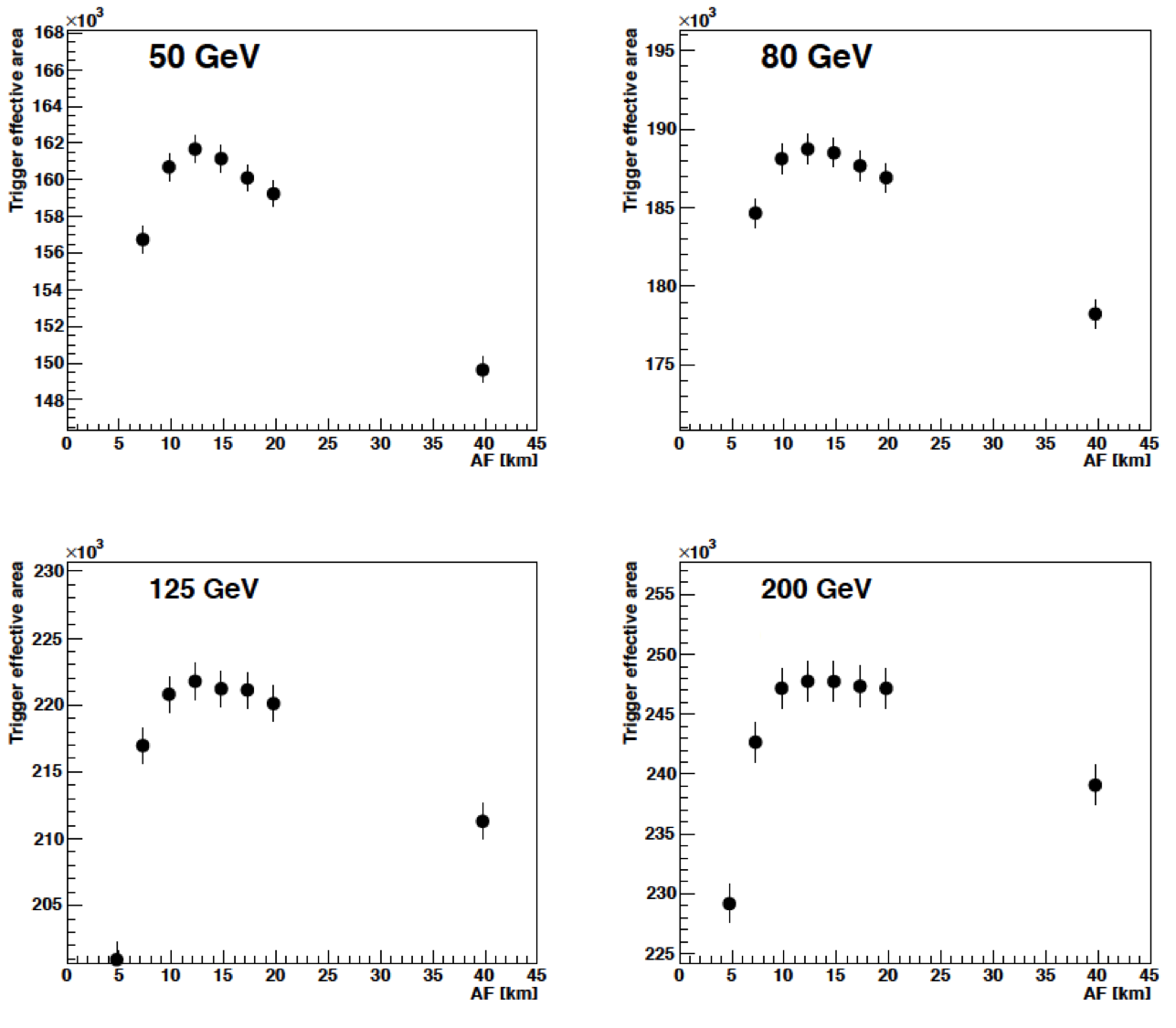}
\caption{\gr\ trigger rate of MC simulations at zenith for different focus altitude and energy. The focus at infinity is shown on the figure at 40~km distance.}
\label{TRgamma}
\end{figure}

Within the simulations package Smash for \hess, the focus of the telescope can be adjusted according to needs. It affects the shape of the images by changing the amount of Cherenkov photons per CT5 camera pixel. Fig.~\ref{EventDisplay50} shows camera images of a simulated \gr\ of 50~GeV seen by CT5 at zenith. The impact of the focus is visible in the larger image for a focus to 5~km and to infinity with respect to the images for a focus at 10~km and 15~km. These latter focus values fall into the typical range of shower development altitude, and generate a correctly focused image. On the other hand, a focus at 5~km and infinity induce an unfocused and thus broader image.

\subsection{Focus impact on the \gr\ trigger rate}
\label{sec:trig}

The work presented here focus mainly on low energies, where the benefits of the focus for mono telescope observation are of prime importance. At higher energies, the multiplicity of the events will reduce the impact of the CT5 focus. Results up to 800 GeV are presented in the following, assuming that most of the events above that threshold will trigger the \hess-I telescopes as well. The low statistics of MC simulations available at low energies prevents to precisely quantify the focus impact below 50 GeV. Only results obtained at zenith are presented, where the focus has the more important impact for this elevation. Similar effects as presented in the following are observed for lower elevation, but with lower amplitude.

Fig~\ref{TRgamma} shows the evolution of the \gr\ trigger rate as a function of focused altitude at zenith and for several fixed energies. A significant increased of the \gr\ trigger rate is visible focusing to typical \gr\ atmospheric shower altitude compared to a focus at infinity. A maximum is observed at $\sim$12.5 km. The gain, compared to infinity, is $\sim$8\% for 50 GeV \grs. This evolution is expected from the image observed on figure~\ref{TRgamma}. When correctly focused, the core of the \gr\ images is more compact and the central pixels of the images are brighter. For a given image, a correct focus on the shower core increases thus the probability for the event to trigger the camera.

\subsection{Comparison between focus at 15~km distance and to infinity}

The mirror alignment was performed during summer 2012 with a focus system at the 0 position which corresponds to a distance between the camera lid surface to the virtual central mirror equivalent to the dish focal distance (36~m modulo the thermal expansion of the steel structure). During the commissioning period, the focus system position was set at 66 mm during observations.  Given the 20.6~mm width of the camera lid, this position of the focus system corresponds to a shift of the entrance plane of the camera to the focal plane of 86.6~mm. This shift translates into a focus at 15~km distance of the telescope. This focus distance was decided according to previous study (see for instance~\cite{Hofmann2001}). A fixed position was decided for the first \hess-II observations. As the camera position is fixed and the depth of field is limited, this means that the focused altitude vary with the observation zenith angle. 

\begin{figure}[!ht]
\centering
\includegraphics[width=0.45\textwidth]{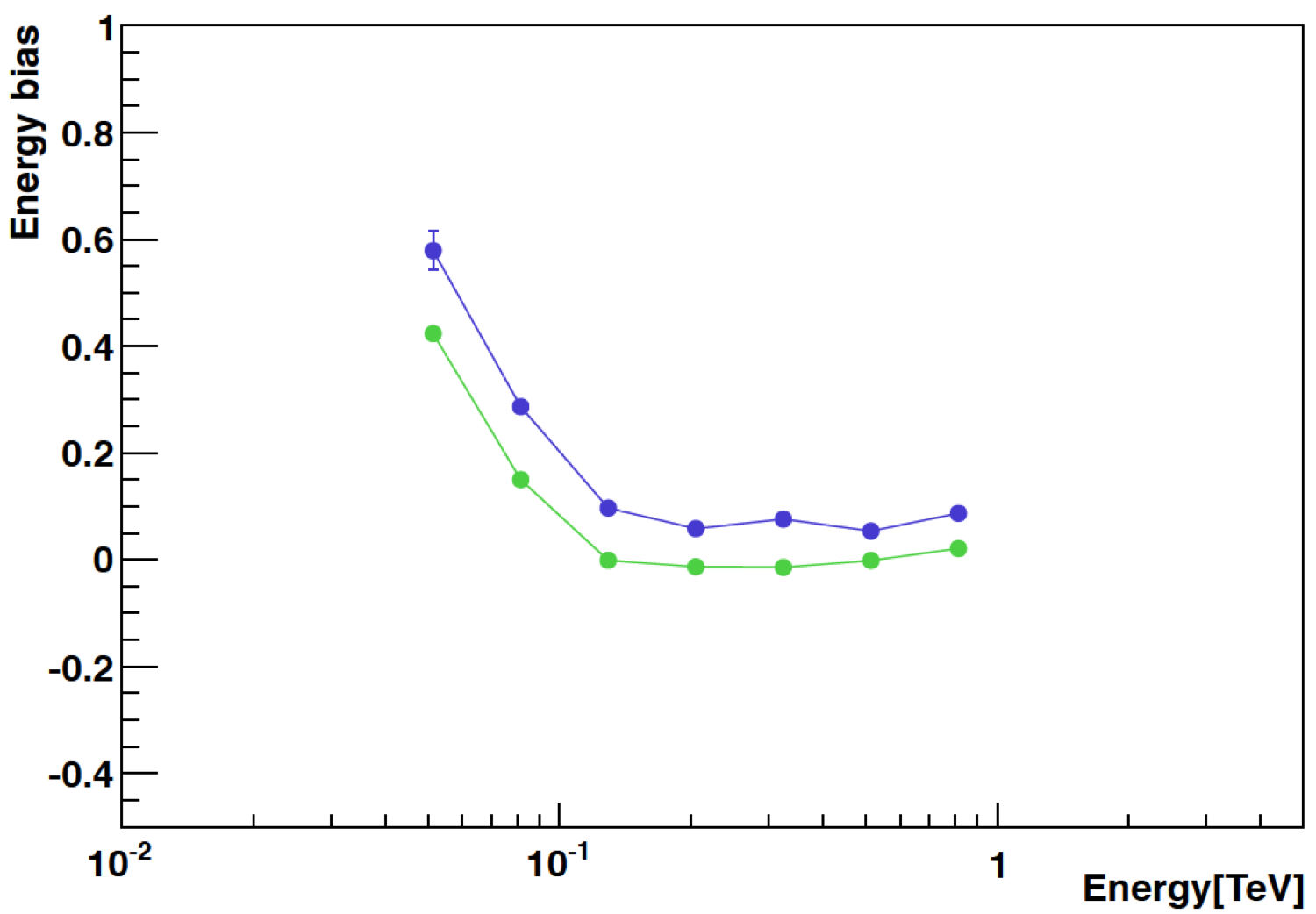}
\includegraphics[width=0.45\textwidth]{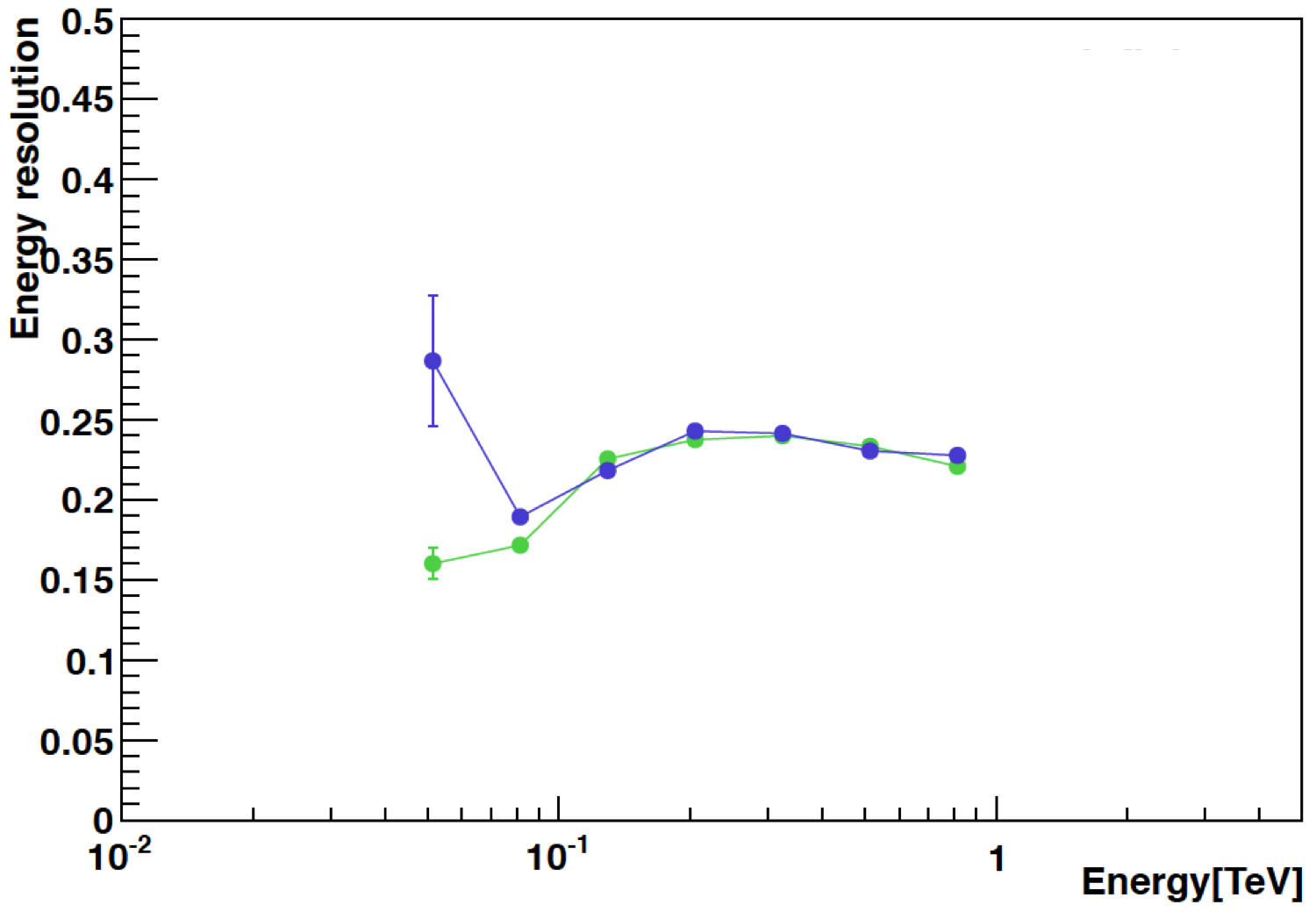} \\
\caption{Left: Evolution of the energy bias as a function of the energy obtained from \gr\ simulation focusing at infinity in green and focusing at 15~km distance in blue. Right: Evolution of the energy resolution as a function of the energy obtained from \gr\ simulation focusing at infinity in green and focusing at 15~km distance in blue.}
\label{Bias}
\end{figure}

The semi analytical {\it model} analysis consists in the comparison pixel by pixel of the obtained images with a semi-analytical model of the \gr\ showers. Therefore it is very sensitive to the shape of the \gr\ images. The focus, which induces a smearing of the images, affects thus the performance of this analysis in several ways.

\begin{figure}[!ht]
\centering
\includegraphics[width=0.6\textwidth]{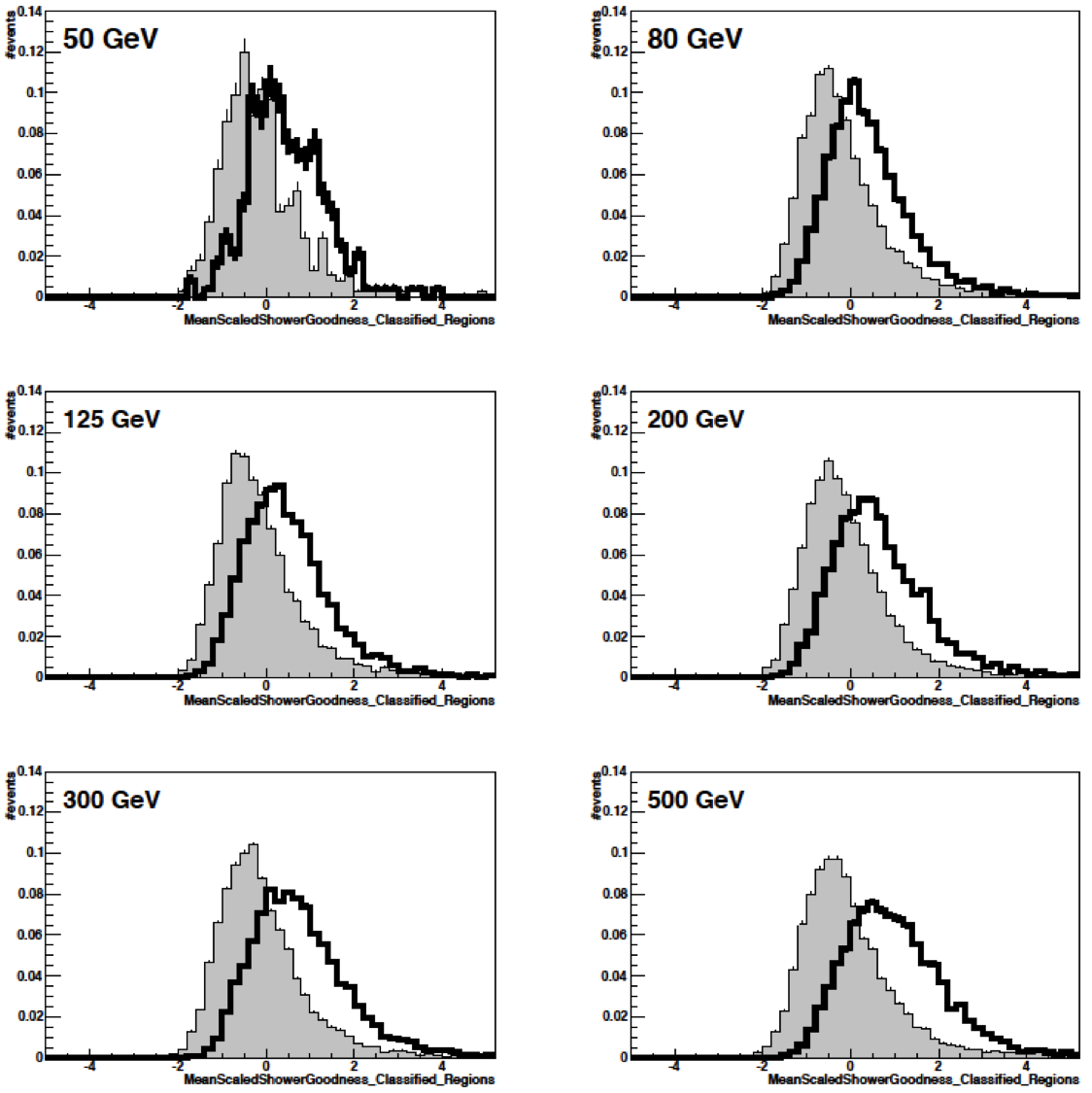} 
\caption{Comparison of the Mean Scaled Shower Goodness distribution obtained from \gr\ Monte Carlo simulations focusing at infinity (filled histogram) and 15~km distance (ComValue - bold line). The panels correspond to several simulated \gr\ at zenith and with energy ranging from 50~GeV to 500~GeV.}
\label{Varcomp_0}
\end{figure}

\begin{itemize}
\item Since {\it model} templates are produced without focus assumption (corresponding to an infinite depth of field), a smeared peaked image will force the fit to preferentially choose a higher energy template. Figure~\ref{Bias} shows the energy bias and resolution of {\it model} mono analysis for a focus at infinity (green) and at 15~km distance (blue). An additional systematic energy bias is introduced by a focus at 15~km distance compared to infinity. Around +10\% of energy bias is introduced at 125GeV at zenith and increases at low energies. The same behavior is observed at higher zenith angles but with a slightly lower amplitude. The energy resolution is not significantly affected by the focus.

\item  The modified shape of the \gr\ images provides a different result of the likelihood ratio test between the fitted {\it model} template and the smeared image. This ends in a different behavior of the main discriminating variable (\emph{MeanScaledShowerGoodness} - hereafter \MSSG ). Figure~\ref{Varcomp_0} shows the distribution of this variable with a focus at infinity, filled in grey, and focusing at 15~km distance in black. A significant shift and broadening with the energy are introduced by the different focus.

\item Others discriminating variables such as \emph{PrimaryDepth} (first interaction atmospheric depth), are good probes of the \gr\ reconstruction power. At low energies, thanks to the more peaked images, the distribution is narrower around 1 (as expected for \grs) revealing the enhanced \gr\ reconstruction power.

\item Control variables such as \emph{MeanScaledBackgroundGoodness} (likelihood of null hypothesis for pixels without expected signal) exhibit normal distributions, showing that the effects of focus are understood.
\end{itemize}

\section{Monte Carlo - Data comparison}

The previous section shows the effects of the modification of \gr\ images due to the focus on the {\it model} reconstruction. The predictions from Monte Carlo simulations have been tested on data. \pks\ has been chosen to test the Monte Carlo simulations because of its important exposure. Its important \gr\ flux allows a substantial ON-OFF data set and therefore a \gr\ sample with low background contamination. The simulation used for the comparison are \gr\ spectrum and the closest configuration from the known spectrum of \pks\ was chosen. A spectrum with an index of -3.0, an offset of 0.5\deg, at zenith, and an azimuth angle of 180\deg,  has been chosen and analyzed using the same selection procedure than the data. 

\begin{figure}[!ht]
\centering
\includegraphics[width=0.35\textwidth]{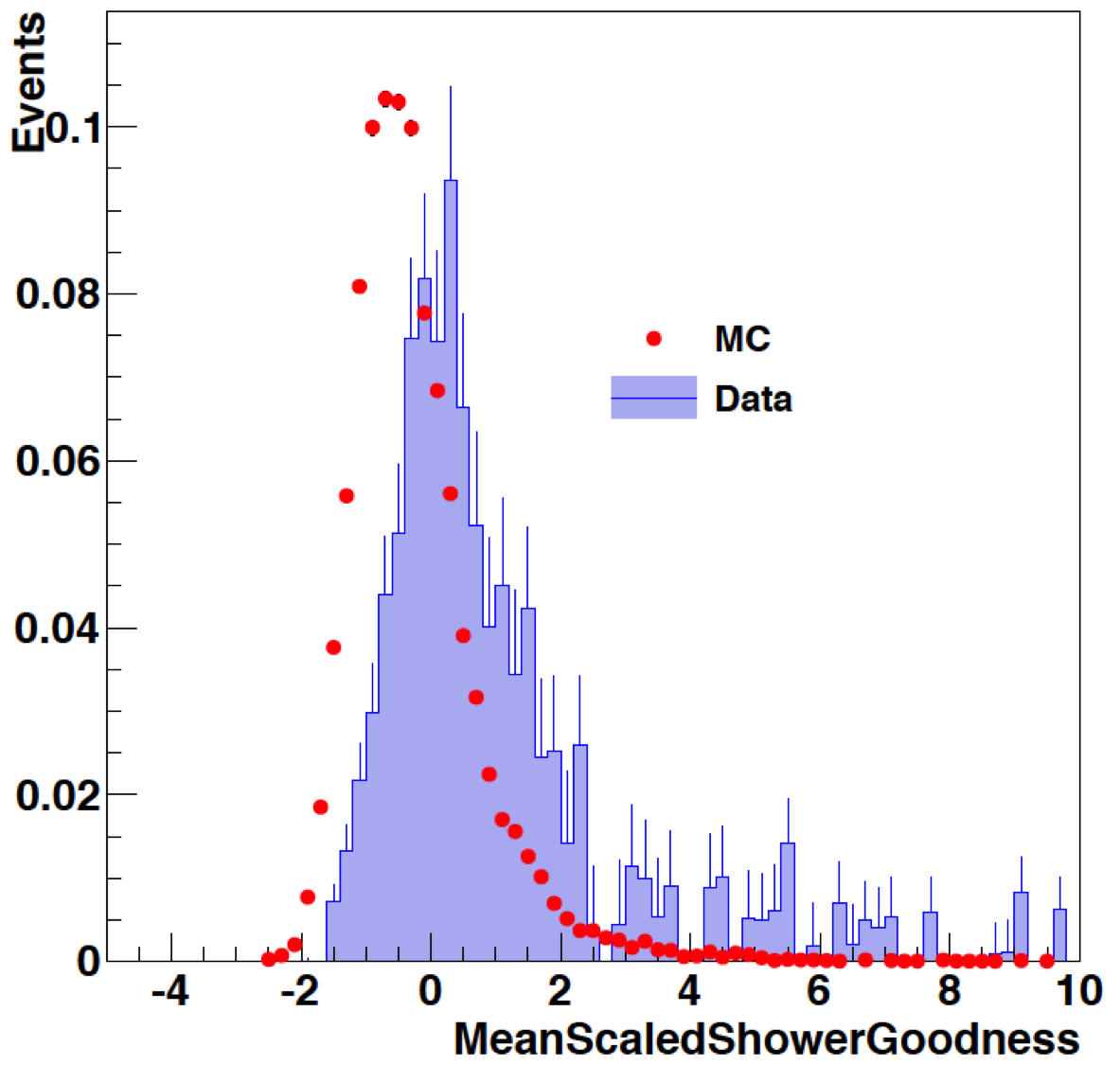}
\includegraphics[width=0.35\textwidth]{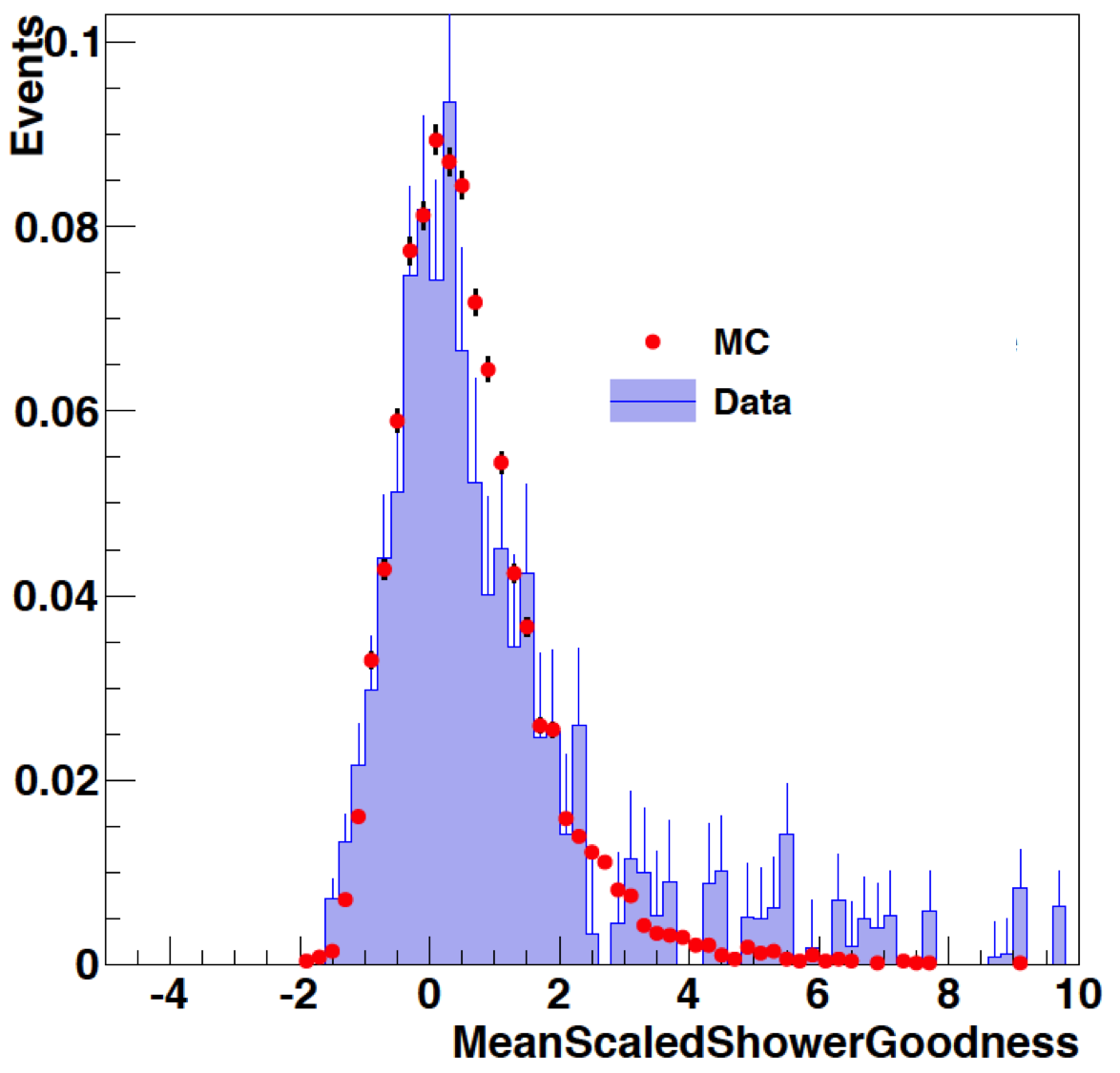}
\caption{Comparison between Monte Carlo simulations following a spectrum with $\Gamma$=-3 (focus at 15~km distance) and data taken on \pks. \emph{Left}:MC with a focus at infinity. \emph{Right}:MC with a focus at 15~km distance. 
\label{Comp_af}}
\end{figure}

Figure~\ref{Comp_af} compares the \MSSG\ distribution obtained on \pks\ data and Monte Carlo simulations. The MC distributions were obtained for \grs\ with a focus at infinity (\emph{left}) and at 15~km distance (\emph{right}). The data have been acquired with a focus at 15~km distance. MC distribution with an focus at 15~km distance matches nicely the data distribution. The shift of the \MSSG\ confirms the focus effects seen in the simulations.

 The very nice agreement between Monte Carlo simulations (with a focus at 15~km as chosen during data acquisition) and Data with the current mono mode analysis, proves that the effects of the focus are understood and fully taken into account.

\section{Instrument response functions (IRFs)}

The focus system is affecting, by changing the shape of the air showers images, each step of \hess\ analysis chain. The \IRFS\ will strongly depend on the focus distance of the telescope. \IRFS\ have been computed, taking the commissioning position of the focus system to establish more realistic description of CT5.

\begin{figure}[!ht]
\centering
\includegraphics[width=0.4\textwidth]{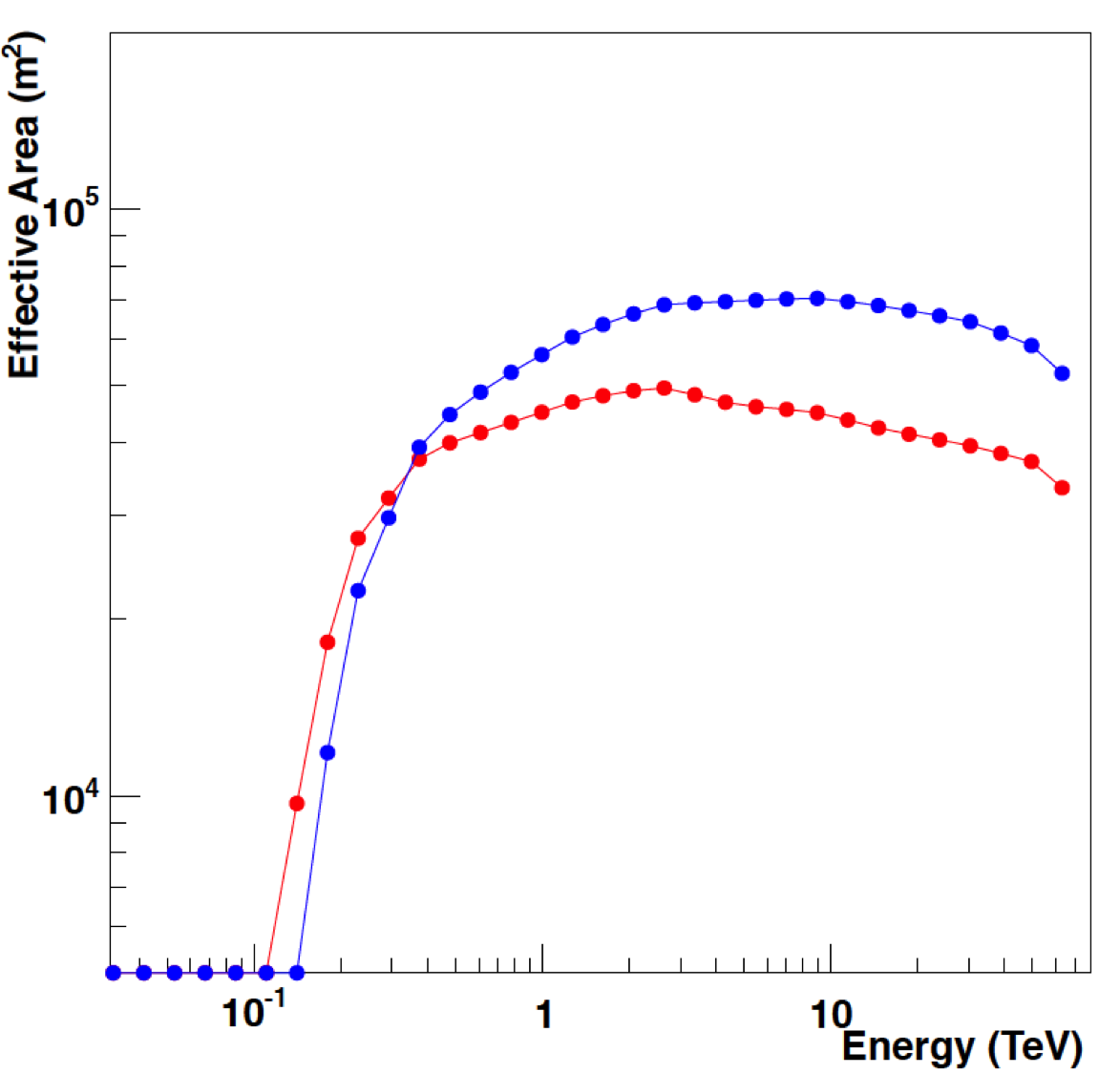}
\includegraphics[width=0.4\textwidth]{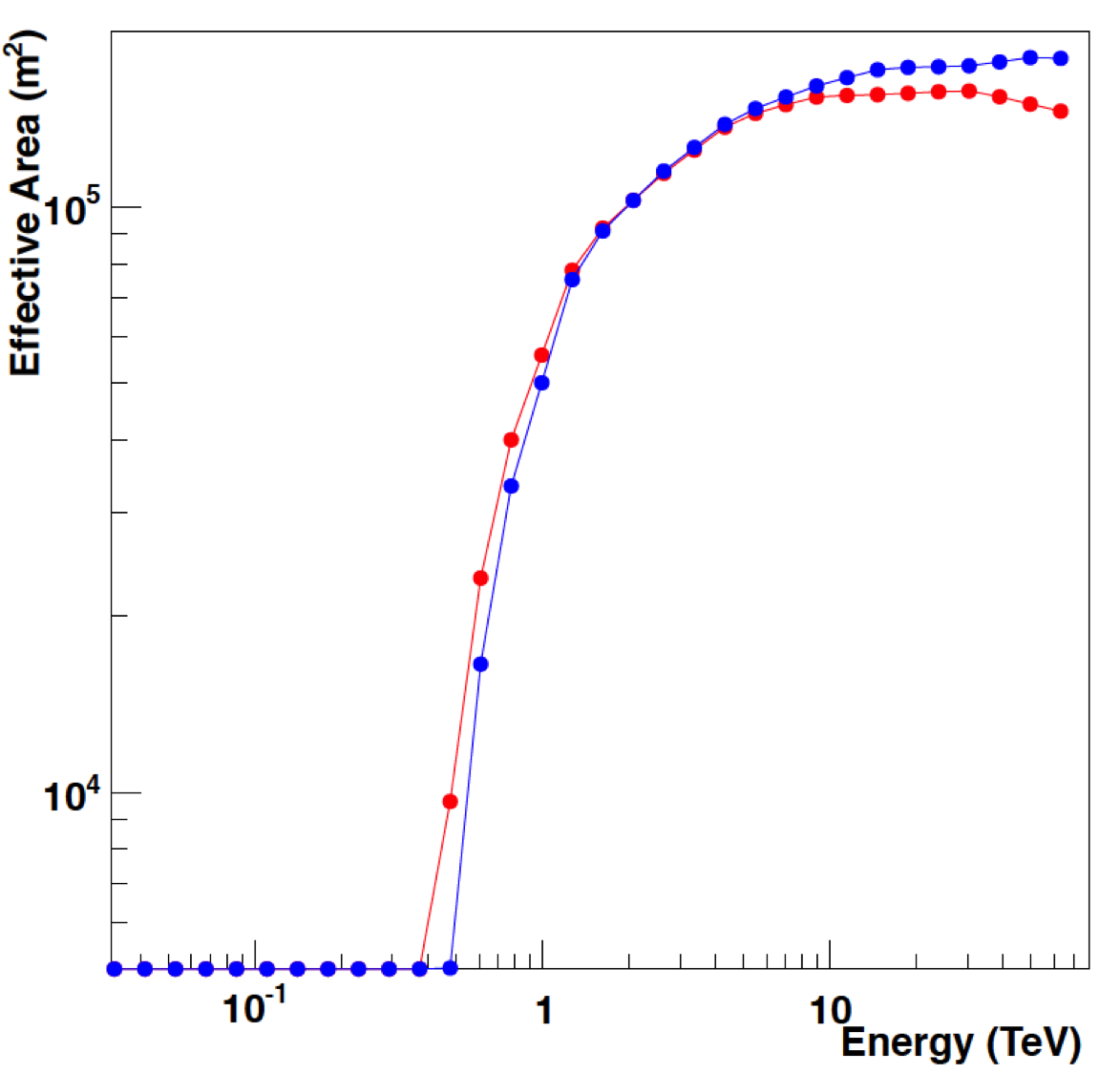} 
\caption{Comparison of \gr\ acceptance focusing at infinity (blue curve) and at 15~km distance (red curve). The left panel is obtained for \gr\ simulations at 18 degrees zenith angle, whereas the right panel is obtained for \gr\ simulations at 50 degrees zenith angle.}
\label{Acceptance}
\end{figure}

Figure~\ref{Acceptance} shows the improvement of the effective area of the \hessii\ {\it model} Mono analysis. The updated acceptance (in red) is increased by a factor of $\sim$9 at $\sim$50~GeV compared to a focus at infinity. The improvement of the acceptance is not only due to the increased trigger rate, only +$\sim$5\% more \grs\ events being expected at 50~GeV (see section~\ref{sec:trig}). Thus, the more focused images strongly enhance the reconstruction power of {\it model} analysis, and thus the event selection cuts.

The acceptance at high energies is reduced by a factor $\sim$2 at $\sim$10TeV. At these energies, the air shower extend more deeply in the atmosphere. The various part of the shower coming from different distance, the focus to a finite distance leads to a smearing varying over the different part of the image. Indeed, the focus altitude at $\sim$15km creates different smearing between the beginning and the end of the shower, whereas the {\it model} analysis assumes an infinite depth of field (all direction or distance perfectly focused). This effect actually mixes with the intrinsic PSF of the mirror facets and the optical aberration coming from the dish mount. To take into account this effect, a constant smearing for all part of the shower is applied and corresponds roughly to the effect expected for showers observed with an infinite focus. The {\it model} reconstruction fit procedure converges on wrong templates and events may be rejected. If the shower image is cut because limited field of view of CT5, which happens more frequently with increasing energy, this effect will be amplified. 

However, this loss can be reduced by introducing the focus smearing in the fit procedure. This implementation is under study and significant improvement is expected. Moreover, at high energies the stereoscopic analyses would compensate the loss due to CT5 only.

\section{Summary}
The fifth \hess\ telescope is equipped with a focus system that allows to displace the camera with respect to the mirrors. The impact of the focus on the analysis of data obtained with this telescope has been studied. It has been shown that applying a correct focus during data taking has a strong impact on data analysis. The focus changes the recorded image morphology, and induces a significant shift in the distribution of all discriminant variables. Focusing close to the altitude of the \gr\ shower maximum allows to maximize the \gr\ acceptance at the energy threshold of the instrument. On the other hand, introducing the focus and more generally the characteristics of the optics (optical aberrations) into the fit procedure of the semi-analytical {\it model} analysis is mandatory to optimize further the analysis performance over the full energy range.
\\
\\

\setstretch{0,5}
{\scriptsize
{\bf Acknowledgement:} The support of the Namibian authorities and of the University of Namibia in facilitating the construction and operation of H.E.S.S. is gratefully acknowledged, as is the support by the German Ministry for Education and Research (BMBF), the Max Planck Society, the German Research Foundation (DFG), the French Ministry for Research, the CNRS-IN2P3 and the Astroparticle Interdisciplinary Programme of the CNRS, the U.K. Science and Technology Facilities Council (STFC), the IPNP of the Charles University, the Czech Science Foundation, the Polish Ministry of Science and Higher Education, the South African Department of Science and Technology and National Research Foundation, and by the University of Namibia. We appreciate the excellent work of the technical support staff in Berlin, Durham, Hamburg, Heidelberg, Palaiseau, Paris, Saclay, and in Namibia in the construction and operation of the equipment.
}

\end{document}